%% file: paper.tex
\documentclass[a4paper,10pt,journal]{IEEEtran}
\IEEEoverridecommandlockouts % enable \thanks

\makeatletter
\newcommand{\AddInputPath}[1]{%
  \ifx\input@path\@undefined
    \def\input@path{#1}
  \else
    \g@addto@macro{\input@path}{#1}
  \fi
}
\makeatother
\AddInputPath{{../}}

\usepackage{etex}

\usepackage[font=footnotesize,caption=false]{subfig} % preload with correct options...

\usepackage{relsize}

\usepackage{array}
\usepackage{booktabs,tabularx}
\usepackage{multirow}
\usepackage[binary-units=true,per-mode=symbol,detect-mode=true]{siunitx}
\usepackage{graphicx}

\usepackage[T1]{fontenc}
\usepackage{textcomp}
\usepackage[utf8]{inputenc}
\usepackage[final]{microtype}
\usepackage{icomma}
\usepackage{xspace}

\usepackage[tbtags]{amsmath}
\usepackage{amssymb,amsfonts,bm}
\usepackage{mathtools} 
\usepackage{dsfont}
\usepackage{mathrsfs}
\usepackage{accents}
\usepackage{empheq}
\usepackage{nccmath}
\usepackage{balance}

\usepackage{color}
\usepackage{calc}
\usepackage{tikz}
\usepackage{pgfplots,pgfplotstable}

\usepackage{pdftexcmds}
\makeatletter
\newcommand{\strequal}[2]{\pdf@strcmp{#1}{#2}==0}
\makeatother

\usepackage[capitalize]{cleveref}
\usepackage{refcount}

\usepackage[inline]{enumitem}
\usepackage{algorithm}
\usepackage{algpseudocode}
\makeatletter
\newcommand{\algmargin}{\the\ALG@thistlm}
\makeatother
\newlength{\whilewidth}
\algdef{SE}[parWHILE]{parWhile}{EndparWhile}[1]
  {\parbox[t]{\dimexpr\linewidth-\algmargin}{%
     \hangindent\whilewidth\strut\algorithmicwhile\ #1\ \algorithmicdo\strut}}{\algorithmicend\ \algorithmicwhile}%
\algnewcommand{\parState}[1]{\State%
  \parbox[t]{\dimexpr\linewidth-\algmargin}{\strut #1\strut}}

\makeatletter
\newcommand\fs@spaceruled{\def\@fs@cfont{\bfseries}\let\@fs@capt\floatc@ruled
  \def\@fs@pre{\vspace{.05in}\hrule height.8pt depth0pt \kern2pt}%
  \def\@fs@post{\kern2pt\hrule\relax}%
  \def\@fs@mid{\kern2pt\hrule\kern2pt}%
  \let\@fs@iftopcapt\iftrue}
\makeatother

\usepackage{glossaries}
\usepackage{ifthen}
\usepackage{cite}
\usepackage{multibib}

\usepackage{comment}
\usepackage{todonotes}
\let\legacytodo\todo
\newcommand{\ruggedtodo}[2][]{\tikzexternaldisable\legacytodo[#1]{#2}\tikzexternalenable}
\renewcommand{\todo}[1]{\ruggedtodo[inline]{#1}}

\input{acronyms}
\glsenableentrycount
\makeglossaries

\usetikzlibrary{positioning}
\usetikzlibrary{calc}
\usetikzlibrary{math}
\usetikzlibrary{fit}
\usetikzlibrary{intersections}
\usetikzlibrary{decorations.pathreplacing}
\usetikzlibrary{decorations.markings}
\usetikzlibrary{3d,angles}

\usetikzlibrary{external}
\pgfkeys{/pgfplots/.cd,
	hk/.style={blue,mark=*},
	snd/.style={red,mark=square*},
	ian/.style={brown!60!black,mark=triangle*},
}

\tikzset{
	antenna/.pic={
		\draw[thick] (0,0) -- ++(120:2mm) -- ++(0:2mm) -- cycle -- (0,-1.5mm);
	}
}

\makeatletter
\newcommand\transformxdimension[1]{
    \pgfmathparse{((#1/\pgfplots@x@veclength)+\pgfplots@data@scale@trafo@SHIFT@x)/10^\pgfplots@data@scale@trafo@EXPONENT@x}
}
\newcommand\transformydimension[1]{
    \pgfmathparse{((#1/\pgfplots@y@veclength)+\pgfplots@data@scale@trafo@SHIFT@y)/10^\pgfplots@data@scale@trafo@EXPONENT@y}
}
\makeatother

\crefname{equation}{}{}
\crefrangeformat{equation}{(#3#1#4)--(#5#2#6)}
\undef\mod
\DeclareMathOperator\mod{mod}

\let\vec\bm

\allowdisplaybreaks[1]

\DeclareSIUnit \dBm {dBm}
\DeclareSIUnit \dBW {dBW}
\DeclareSIUnit \bpcu {bpcu}

\usepackage{etoolbox}
\AtBeginEnvironment{equation}{\small}
\AtEndEnvironment{equation}{\normalfont}
\AtBeginEnvironment{gather}{\small}
\AtEndEnvironment{gather}{\normalfont}
\AtBeginEnvironment{split}{\small}
\AtEndEnvironment{split}{\normalfont}
\AtBeginEnvironment{equation*}{\small}
\AtEndEnvironment{equation*}{\normalfont}
\AtBeginEnvironment{align}{\small}
\AtEndEnvironment{align}{\normalfont}
\AtBeginEnvironment{align*}{\small}
\AtEndEnvironment{align*}{\normalfont}
\AtBeginEnvironment{alignat}{\small}
\AtEndEnvironment{alignat}{\normalfont}
\AtBeginEnvironment{flalign}{\small}
\AtEndEnvironment{flalign}{\normalfont}
\AtBeginEnvironment{multline}{\small}
\AtEndEnvironment{multline}{\normalfont}
\AtBeginEnvironment{multline*}{\small}
\AtEndEnvironment{multline*}{\normalfont}

\DeclareFontFamily{U}{mathx}{\hyphenchar\font45}
\DeclareFontShape{U}{mathx}{m}{n}{
      <5> <6> <7> <8> <9> <10>
      <10.95> <12> <14.4> <17.28> <20.74> <24.88>
      mathx10
      }{}
\DeclareSymbolFont{mathx}{U}{mathx}{m}{n}
\DeclareMathSymbol{\bigtimes}{1}{mathx}{"91}

\newtheorem{theorem}{Theorem}

\pgfplotscreateplotcyclelist{default}{%
	blue,mark=*\\%
	red,mark=star\\%
	teal,mark=square*\\%
	brown!60!black,mark=otimes*\\%
}

\pgfplotscreateplotcyclelist{long}{%
	red,solid,mark=*\\%
	blue,solid,mark=square*\\%
	cyan,densely dashed,every mark/.append style={solid},mark=triangle*\\%
	teal,densely dashed,every mark/.append style={solid},mark=star\\%
	brown!60!black,densely dashed,every mark/.append style={solid},mark=diamond*\\%
}

\newcommand\Tx{\mathrm{T}}
\newcommand\Rx{\mathrm{R}}
\newcommand\s{\mathrm{s}}
\renewcommand\d{\mathrm{d}}

\makeatletter
\newif\ifhbonecolumn
\@ifclasswith{IEEEtran}{onecolumn}{\hbonecolumntrue}{\hbonecolumnfalse}
\makeatother

\begin{document}
\bstctlcite{IEEEexample:BSTcontrol}
%\title{\pp{Dynamic Digital Beamforming towards a Satellite using Intelligent Reflecting Surfaces (IRS)}}
%\title{Intelligent Reflecting Surface (IRS) Operation\\ under Predictable Receiver Mobility:\\ A Continuous Time Propagation Model}
\title{Intelligent Reflecting Surface Operation\\ under Predictable Receiver Mobility:\\ A Continuous Time Propagation Model}

\author{Bho~Matthiesen,~\IEEEmembership{Member,~IEEE},
		Emil~Björnson,~\IEEEmembership{Senior~Member,~IEEE},
		Elisabeth~De~Carvalho,~\IEEEmembership{Senior~Member,~IEEE},
		and~Petar~Popovski,~\IEEEmembership{Fellow,~IEEE}%
		\thanks{
			B.~Matthiesen is with the Department of Communications Engineering, University of Bremen, 28359 Bremen, Germany (e-mail: matthiesen@uni-bremen.de).
			E.~Björnson is with the Department of Electrical Engineering (ISY), Linköping University, 58183 Linköping, Sweden (e-mail: emil.bjornson@liu.se).
			E.~De~Carvalho and P.~Popovski are with the Department of Electronic Systems, Aalborg University, 9100 Aalborg, Denmark (e-mail: \{edc, petarp\}@es.aau.dk). P.~Popovski is also holder of the U Bremen Excellence Chair in the Department of Communications Engineering, University of Bremen, 28359 Bremen, Germany.
		}%
		\thanks{
			This work is supported in part by the German Research Foundation (DFG)
			under Germany's Excellence Strategy (EXC 2077 at University of Bremen, University Allowance), in part by ELLIIT and in part by the Danish Council for Independent Research DFF-701700271.
		}
	}

\maketitle
%\tikzset{external/force remake}
%\tikzset{external/export=false}

\begin{abstract}
	The operation of an intelligent reflecting surface (IRS) under predictable receiver mobility is investigated.
	We develop a continuous time system model for multipath channels and discuss the optimal IRS configuration with respect to received power, Doppler spread, and delay spread. It is shown that the received power can be maximized without adding Doppler spread to the system. In a numerical case study, we show that an IRS having the size of just two large billboards can improve the link budget of ground to Low Earth Orbit (LEO) satellite links by up to 6\,dB.
	It also adds a second, almost equivalently strong, communication path that improves the link reliability.
\end{abstract}
\glsresetall


\section{Introduction}
\cGlspl{irs} are an emerging technology that enables tunable anomalous scattering of incident electromagnetic waves \cite{Liaskos2018,DiRenzo2019,Basar2019,Wu2020}.
This permits active control of the propagation environment and introduces an additional optimization dimension to wireless communication networks.
A main use case is range extension and 
recent work focuses mostly on maximizing the received power over the \cgls{irs} path \cite{Wu2019,Bjornson2020,Tang2019}. Instead, this paper considers multipath propagation with predictive receiver mobility and evaluates the implications of adding an \cgls{irs} to a \cgls{los} communication scenario.
This calls for the development of a continuous time model that has, to the best of the authors' knowledge, not appeared in previous works. We uncover fundamental phenomena that are not visible in the standard discrete time models.

Our motivation to consider this setup is uplink transmission of an \cgls{iot} device to a satellite in \cgls{leo}.
Since the satellite's orbit is known a~priori, its position is completely predictable at all times. We show that this information can be used at an \cgls{irs} to optimize the received signal at the satellite in terms of \cgls{snr}, Doppler spread, or delay spread. In particular, the \cgls{snr} can be maximized while simultaneously compensating Doppler spreading entirely and keeping excess delay spread due to \cgls{irs} operation within one carrier signal period. Numerical results show an \cgls{snr} gain of \SIrange{3}{6}{\dB} for an \cgls{irs} the size of two large billboards that can be achieved without introducing Doppler spread into the system. It is shown that this gain cannot be achieved by using a simple reflector of the same size and, hence, is due to the proposed optimal choice of phase shifts at the \cgls{irs}. To the best of the authors knowledge, this is the first work to optimize \cgls{irs} operation with predictive mobility compensation. Other application scenarios of our results include communication with vehicles on predictable paths, e.g., a car on a highway, a train, or an airplane. In the following, we first consider a generic mobility model and then, in \cref{sec:numeval}, apply our results to the outlined satellite communication scenario.

\paragraph*{Notation}
Vectors are typeset in bold face. Euclidean points are defined as $\vec p = (x, y, z)^T$. The functions $\Re\{\cdot\}$, $\lceil \cdot \rceil$, and $\lfloor \cdot \rfloor$ give the real value, the ceiling, and the floor of their argument, respectively. Further, $\mod(x, y)$ is the remainder of the division of $x$ by $y$, norms are $L^2$, $j$ is the imaginary unit, $e$ is Euler's number, $\mathds Z$ and $\mathds N$ are the sets of integers and nonnegative numbers, respectively.

\begin{comment}
\pp{Key points:
\begin{itemize}
    \item IRS can be used to create multipath towards a satellite
    \item Predictable satellite movement allows for adaptive dynamic change of the phases in IRS in order to optimize the received signal in terms of SNR, Doppler spread or delay spread.
    \item Conflicting objectives, we show that SNR maximization and Doppler compensation are compatible.
    \item To the best of our knowledge, one of the first works to investigate IRS for scenarios with predictive mobility compensation.
    \item In practice, IRS should also be physically rotated, just as a satellite dish, in order to improve the gains.
    \item Comparison to direct path-only as well as billboard with random phases. 
\end{itemize}}
\end{comment}

\section{System Model}
\begin{figure}
	\centering
	\includegraphics{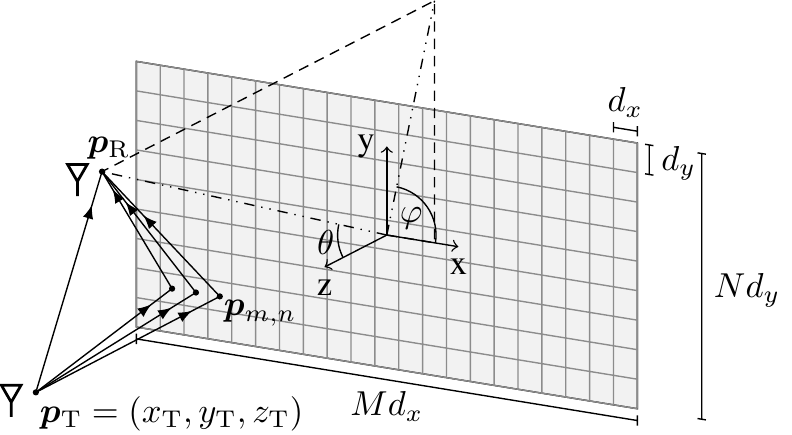}
	\caption{Illustration of system model.}
	\vspace{-2ex}
	\label{fig:sysmod}
\end{figure}

We consider an \cgls{irs} in the $xy$-plane of a Cartesian coordinate system with its geometric center at the origin. It consists of $M$ columns and $N$ rows of reflecting elements placed on a rectangular grid spaced $d_x$ and $d_y$ apart. The dimensions $d_x$ and $d_y$ of each \cgls{irs} element are usually within the range of $\frac{\lambda_c}{10}$ and $\frac{\lambda_c}{5}$ \cite{Ozdogan2020a}, where $\lambda_c$ is the carrier wavelength. The center of element $(m, n)$, $m\in\mathcal G(M)$ and $n\in\mathcal G(N)$ with
\begin{equation}
	\mathcal G(M) = \left\{ \mod(M+1, 2) - \left\lfloor \frac{M}{2} \right\rfloor, \dots, \left\lfloor \frac{M}{2} \right\rfloor \right\},
\end{equation}
is $\vec p_{m, n} = (g(m, d_x, M), g(n, d_y, N), 0)$ with
$g(m, d_x, M) = m d_x - 0.5 d_x \mod(M+1, 2)$.
Each element has antenna gain $G(\theta_{m, n}, \varphi_{m, n})$ with polar angle $\theta_{m, n} \in [0, \pi]$ and azimuth angle $\varphi_{m, n} \in [0, 2\pi]$ as indicated in \cref{fig:sysmod}. As the \cgls{irs} can only receive power from one side, we assume that $G(\theta_{m, n}, \varphi_{m, n}) = 0$ for $\theta_{m, n} \in[\frac{\pi}{2}, \pi]$.
The single-antenna transmitter and single-antenna receiver have antenna gains $G_\Tx(\theta_\Tx, \varphi_\Rx)$ and $G_\Rx(\theta_\Rx, \varphi_\Rx)$, with $\theta_\Tx, \theta_\Rx\in[0, \pi]$ and $\varphi_\Tx, \varphi_\Rx\in[0, 2\pi]$, respectively, and are located at $\vec p_t$ and $\vec p_r(t)$. The \cgls{irs} and transmitter have fixed positions, while the receiver moves with time $t$.
We do not consider polarization losses and coupling between the \cgls{irs} elements. We also assume that all transmissions take place in the far-field. The analysis can potentially be extended to other cases and we refer to \cite{Williams2019,Bjornson2020} for more information on coupling and near-field analysis.

We consider \cgls{los} transmission between the transmitter and receiver, both over the direct path and over the \cgls{irs}.%
\footnote{Local scattering around the transmitter is neglected for simplicity. We expect its effects to simply propagate through the \cgls{irs} given that the optimization in \cref{sec:opt} is done for the \cgls{los} path. As this path usually has the maximum power, this is a reasonable assumption.}
The transmitter emits a passband signal $x_p(t) = \sqrt{2}\, \Re\{ x(t) e^{j 2 \pi f_c t} \}$
with carrier frequency $f_c$. It is created from the complex-baseband signal
$x(t) = x_i(t) + j x_q(t)$.
The components $x_i(t)$ and $x_q(t)$ are baseband bandlimited to $B/2$. In the absence of other propagation paths, the signal observed at an infinitesimal point at position $\vec p$ and time $t$ is
\begin{equation}
	\frac{\sqrt{G_\Tx(\theta_\Tx, \varphi_\Tx)}}{ \sqrt{4\pi} \Vert \vec p - \vec p_\Tx \Vert} x_p\!\left(t - \frac{\Vert \vec p - \vec p_\Tx \Vert}{c_0}\right),
\end{equation}
where the angles $\theta_\Tx$ and $\varphi_\Tx$ are computed between $\vec p_\Tx$ and $\vec p$ as $\theta_\Tx = \theta(\vec p_\Tx, \vec p)$ and $\varphi_\Tx = \varphi(\vec p_\Tx, \vec p)$ with
$\theta(\vec p_1, \vec p_2) =  \arccos \left( \frac{z_2 - z_1}{\Vert \vec p_2 - \vec p_1 \Vert} \right)$ and 
$\varphi(\vec p_1, \vec p_2) = \arctan\left( \frac{y_2 - y_1}{x_2 - x_1} \right)$,
where the inverse tangent is defined such that it takes into account the correct quadrant for $(x_2 - x_1, y_2-y_1)$.

Adding the receive antenna, which has effective area $G_\Rx(\theta(\vec p_\Rx(t), \vec p_\Tx), \varphi(\vec p_\Rx(t), \vec p_\Tx)) \lambda_c^2/(4\pi)$, the observed signal (in the presence of only the direct \cgls{los} path) at the receiver is
\begin{equation} \label{eq:rxsignal}
	\frac{\lambda_c \sqrt{G_\Tx^\Rx(t)\; G_\Rx^\Tx(t)}}{ 4\pi \Vert \vec p_\Rx(t) - \vec p_\Tx \Vert} x_p\!\left(t - \frac{\Vert \vec p_\Rx(t) - \vec p_\Tx \Vert}{c_0}\right)
\end{equation}
where $G_\Tx^\Rx(t)$ is the antenna gain of the transmit antenna in the direction of the receiver and $G_\Rx^\Tx(t)$ is the antenna gain of the receiver in the direction of the transmitter. These terms are computed as
\begin{subequations} \label{eq:antgains}
	\begin{align}
		G_\Tx^\Rx(t) &= G_\Tx(\theta(\vec p_\Tx, \vec p_\Rx(t)), \varphi(\vec p_\Tx, \vec p_\Rx(t))),\\
		G_\Rx^\Tx(t) &= G_\Rx(\theta(\vec p_\Rx(t), \vec p_\Tx), \varphi(\vec p_\Rx(t), \vec p_\Tx)),
	\end{align}
\end{subequations}
with $\theta$ and $\varphi$ defined as above.

%Similarly, the signal observed at time $t$ by the $(m, n)$th \cgls{irs} element is
%\begin{equation}
%\frac{\lambda_c \sqrt{G_\Tx^{m, n}\; G_{m, n}^\Tx}}{4\pi \Vert \vec p_{m, n} - \vec p_\Tx \Vert} x_p\!\left(t - \frac{\Vert \vec p_{m, n} - \vec p_\Tx \Vert}{c_0}\right)
%\end{equation}
%where the antenna gains \chg{$G_\Tx^{m, n}$ and $G_{m, n}^\Tx$ are computed in the same way as \cref{eq:antgains}.}
%\begin{align}
%	G_\Tx^{m, n} &= G_\Tx(\theta(\vec p_\Tx, \vec p_{m, n}), \varphi(\vec p_\Tx, \vec p_{m, n})),\\
%	G_{m, n}^\Tx &= G(\theta(\vec p_{m, n}, \vec p_\Tx), \varphi(\vec p_{m, n}, \vec p_\Tx)).
%\end{align}
The signal observed at time $t$ by the $(m, n)$th \cgls{irs} element is similar to \cref{eq:rxsignal} and
%The observed signal
is time-delayed by $\frac{\phi_{m, n}(t)}{2\pi f_c} \ge 0$ before being scattered. This time delay is controllable in an \cgls{irs} and leads to a phase shift of $\phi_{m, n}(t)$ in the emitted signal. More precisely, the re-emitted signal is
\begin{equation}
	\sqrt{\mu} %\sqrt{G(\theta_{m, n}', \varphi_{m, n}')}
	\frac{\lambda_c \sqrt{G_\Tx^{m, n}\; G_{m, n}^\Tx}}{4\pi \Vert \vec p_{m, n} - \vec p_\Tx \Vert} x_p\!\left(t - \frac{\Vert \vec p_{m, n} - \vec p_\Tx \Vert}{c_0} - \frac{\phi_{m, n}(t)}{2\pi f_c} \right),
\end{equation}
where the antenna gains $G_\Tx^{m, n}$ and $G_{m, n}^\Tx$ are computed in the same way as \cref{eq:antgains} and
where $\mu \in [0,1]$ determines the fraction of the incident energy that is scattered.
The signal that reaches the receiver over this propagation path is
\ifhbonecolumn
\begin{equation}
	\sqrt{\mu} \frac{\lambda_c \sqrt{G_{m, n}^\Rx(t)\, G_\Rx^{m, n}(t)}}{4\pi \Vert \vec p_\Rx(t) - \vec p_{m, n} \Vert} \frac{\lambda_c \sqrt{G_\Tx^{m, n}\; G_{m, n}^\Tx}}{4\pi \Vert \vec p_{m, n} - \vec p_\Tx \Vert}
	\
	x_p\!\left(t - \frac{\Vert \vec p_{m, n} - \vec p_\Tx \Vert}{c_0} - \frac{\phi_{m, n}(t)}{2\pi f_c} - \frac{\Vert \vec p_\Rx(t) - \vec p_{m, n} \Vert}{c_0} \right)
\end{equation}
\else
\begin{multline}
	\sqrt{\mu} \frac{\lambda_c \sqrt{G_{m, n}^\Rx(t)\, G_\Rx^{m, n}(t)}}{4\pi \Vert \vec p_\Rx(t) - \vec p_{m, n} \Vert} \frac{\lambda_c \sqrt{G_\Tx^{m, n}\; G_{m, n}^\Tx}}{4\pi \Vert \vec p_{m, n} - \vec p_\Tx \Vert}
	\\
	x_p\!\left(t - \frac{\Vert \vec p_{m, n} - \vec p_\Tx \Vert}{c_0} - \frac{\phi_{m, n}(t)}{2\pi f_c} - \frac{\Vert \vec p_\Rx(t) - \vec p_{m, n} \Vert}{c_0} \right)
\end{multline}
\fi
with antenna gains $G_{m, n}^\Rx(t)$ and $G_\Rx^{m, n}(t)$ computetd as in \cref{eq:antgains}.
%\begin{align}
%	G_{m, n}^\Rx(t) &= G(\theta(\vec p_{m, n}, \vec p_\Rx(t)), \varphi(\vec p_{m, n}, \vec p_\Rx(t))),
%	\\
%	G_\Rx^{m, n}(t) &= G_\Rx(\theta(\vec p_\Rx(t), \vec p_{m, n}), \varphi(\vec p_\Rx(t), \vec p_{m, n})).
%\end{align}

Assuming the propagation channel consists of the direct path and the scattered paths via the \cgls{irs}, we obtain the received passband signal as
\ifhbonecolumn
\begin{equation}
	y_p(t) = A_0(t) x_p(t - \tau_0(t))
	+ \sum_{m, n} A_{m, n}(t) x_p\!\left(t - \tau_{m, n}(t) - \frac{\phi_{m, n}(t)}{2\pi f_c}\right) + n_p(t)
\end{equation}
\else
\begin{align}
	\MoveEqLeft y_p(t) = A_0(t) x_p(t - \tau_0(t)) \notag \\
	&+ \sum_{m, n} A_{m, n}(t) x_p\!\left(t - \tau_{m, n}(t) - \frac{\phi_{m, n}(t)}{2\pi f_c}\right) + n_p(t)
\end{align}
\fi
with $n_p(t)$ being a \cgls{wgn} process with power spectral density $N_0$.
The amplitude gains and delays are
\begin{align}
	A_0(t) &= \frac{\lambda_c \sqrt{G_\Tx^\Rx(t)\; G_\Rx^\Tx(t)}}{ 4\pi \Vert \vec p_\Rx(t) - \vec p_\Tx \Vert},
	&
	\tau_0(t) &= \frac{\Vert \vec p_\Rx(t) - \vec p_\Tx \Vert}{c_0},
\end{align}
for the direct path and
\begin{align}
	A_{m, n}(t) &= \sqrt{\mu} \frac{\lambda_c^2}{16\pi^2} \frac{ \sqrt{G_{m, n}^\Rx(t)\, G_\Rx^{m, n}(t)\, G_\Tx^{m, n}\; G_{m, n}^\Tx}}{\Vert \vec p_\Rx(t) - \vec p_{m, n} \Vert \Vert \vec p_{m, n} - \vec p_\Tx \Vert},
	\\
	\tau_{m, n}(t) &= \frac{\Vert \vec p_{m, n} - \vec p_\Tx \Vert + \Vert \vec p_\Rx(t) - \vec p_{m, n} \Vert}{c_0}
\end{align}
for the path over the $(m, n)$th \cgls{irs} element. Observe that the tunable delay $\frac{\phi_{m, n}(t)}{2\pi f_c}$ of the \cgls{irs} is not included in $\tau_{m, n}(t)$ as it can be directly controlled to optimize the propagation environment.
The corresponding complex baseband signal is
\ifhbonecolumn
\begin{equation} \label{eq:bbrecv}
y(t) = A_0(t) e^{-j 2\pi f_c \tau_0(t)} x(t - \tau_0(t)) + \sum_{m, n} A_{m, n}(t) e^{-j 2\pi f_c \tau_{m, n}(t) - j \phi_{m, n}(t)}  x\!\left(t - \tau_{m, n}(t) - \frac{\phi_{m, n}(t)}{2\pi f_c}\right) + n(t).
\end{equation}
\else
\begin{multline} \label{eq:bbrecv}
y(t) = A_0(t) e^{-j 2\pi f_c \tau_0(t)} x(t - \tau_0(t)) \\+ \sum_{m, n} A_{m, n}(t) e^{-j 2\pi f_c \tau_{m, n}(t) - j \phi_{m, n}(t)} \\ x\!\left(t - \tau_{m, n}(t) - \frac{\phi_{m, n}(t)}{2\pi f_c}\right) + n(t).
\end{multline}
\fi
It can be observed that the complex pseudo-baseband channel response\footnote{The complex baseband channel response is bandlimited and obtained by low-pass filtering \cref{eq:bbchan}.} of the $(m, n)$th \cgls{irs} path is
\begin{equation} \label{eq:bbchan}
	A_{m, n}(t) e^{-j 2\pi f_c \tau_{m, n}(t)} e^{- j \phi_{m, n}(t)}.
\end{equation}
This matches with the usual narrow-band model of \cgls{irs} communication, e.g., used in \cite{Basar2019,Wu2019,Tang2019,Ozdogan2020a} and others, where the \cgls{irs} operation is represented by a multiplicative coefficient $\sqrt{\mu} e^{j \gamma_{m, n}}$. Observe that causality requires $\gamma_{m, n}$ to be negative as opposed to the usual assumption of $\gamma_{m, n}\ge0$.

%\todo[inline]{corresponding baseband noise $n(t)$ satisfies $n_p(t) = \sqrt{2} \Re\{ n(t) \}$ and is circularly-symmetric complex \cgls{wgn} process with power spectral density $N_0$.}

\section{Optimizing the \cgls{irs} Operation} \label{sec:opt}

In the considered setup, the sole purpose of the \cgls{irs} is to improve the channel between the transmitter and the receiver.
In the described \cgls{los} propagation environment, the direct channel is subject to a transmission delay and a Doppler shift. Introducing the \cgls{irs} turns this channel into a multi-path environment with the usual positive effects of increased received power and transmission diversity. It also potentially introduces delay spread and Doppler spread which necessitates more complex receivers and might degrade the performance.
As the adjustable phase shifts of the \cgls{irs} can be configured in almost arbitrary ways, any of these multi-path effects can be either amplified or attenuated. We will first independently optimize the phase shifts with respect to maximum received power, minimum Doppler spread, and minimum delay spread. Then, we discuss trade-offs between these scalar solutions and discuss the joint optimization with respect to these metrics, using the multi-objective optimization methodology \cite{Zadeh1963,Bjornson2014}. %show how the received power can be maximized while keeping negative spreading effects to a minimum.

\subsection{Received Power}
Assume the transmitter is sending a constant signal with power $P_\Tx$.
Then, the average received power in a time interval of length $2T$ centered around time instance $t_0$ is
\begin{equation}
	\frac{1}{2T} \int_{t_0 - T}^{t_0 + T} \vert y(t) - n(t) \vert^2 dt
	=
	\frac{1}{2T} \int_{t_0 - T}^{t_0 + T} P_{\Rx}(t) dt,
\end{equation}
where $P_{\Rx}(t)$ is the instantaneous receive power
\begin{equation}
	P_\Tx \Big| A_0(t) e^{-j 2\pi f_c \tau_0(t)} + \sum_{m, n} A_{m, n}(t) e^{-j 2\pi f_c \tau_{m, n}(t) - j \phi_{m, n}(t)} \Big\vert^2.
\end{equation}
Due to the monotonicity of integration \cite[Thm~12.4]{Bauer2001}, the average received power is maximized if $P_\Rx(t)$ is maximized at every $t$. It follows from the triangle inequality that this requires all the terms to have the same phase
\cite{Wu2019,Ozdogan2020a}, i.e.,
\begin{equation}
	2\pi f_c \tau_0(t) = 2\pi f_c \tau_{m, n}(t) + \phi_{m, n}(t) - 2\pi k_{m, n}(t)
\end{equation}
for all $m, n$ and arbitrary $k_{m, n}(t)\in\mathds Z$, where $k_{m, n}(t)$ are additional full carrier signal period delays that do not affect the received power. Hence, choosing
\begin{equation} \label{eq:Popt}
	\phi_{m, n}(t) = 2\pi f_c (\tau_0(t) - \tau_{m, n}(t)) + 2\pi k_{m, n}(t)
\end{equation}
maximizes the receive \cgls{snr}.
This resembles the result in previous works \cite{Wu2019,Ozdogan2020a} where it is obtained from discrete time models.
The causality requirement $\phi_{m, n}(t) \ge 0$ implies
\begin{equation} \label{eq:kcaus}
	k_{m, n}(t) \ge f_c (\tau_{m, n}(t) - \tau_0(t)).
\end{equation}
It follows from the triangle inequality that
\begin{align}
	\tau_{m, n}(t)
	%&= \frac{\Vert \vec p_{m, n} - \vec p_\Tx \Vert + \Vert \vec p_\Rx(t) - \vec p_{m, n} \Vert}{c_0}
	%\notag\\
	&\ge \frac{\Vert \vec p_{m, n} - \vec p_\Tx + \vec p_\Rx(t) - \vec p_{m, n} \Vert}{c_0}
	%= \frac{\Vert \vec p_\Rx(t) - \vec p_\Tx \Vert}{c_0}
	= \tau_0(t)
	\label{eq:delayineq}
\end{align}%
for all $m, n$. Hence, $k_{m, n}(t)\in\mathds N$ and, unless $\tau_{m, n}(t) = \tau_0(t)$, $k_{m, n}(t) \ge 1$.

\subsection{Doppler Spread}
The Doppler spread is the maximum difference in instantaneous frequency over all significant propagation paths, i.e.,
$D_\s(t) = \max\{ D_{\s,0}(t), D_\mathrm{s,\cgls{irs}}(t) \}$
where
\begin{equation}
	D_{\s,0}(t) = f_c \max_{m, n} \left\vert \frac{d}{dt} \left( \tau_{m, n}(t) + \frac{\phi_{m, n}(t)}{2\pi f_c} \right) - \frac{d}{dt} \tau_0(t) \right\vert
\end{equation}
is the Doppler spread between the direct path and the \cgls{irs},
and
\ifhbonecolumn
\begin{equation}
	D_\mathrm{s,\cgls{irs}}(t) = f_c \max_{m, n,m',n'} \bigg\vert \frac{d}{dt} \left( \tau_{m, n}(t) + \frac{\phi_{m, n}(t)}{2\pi f_c} \right)
	- \frac{d}{dt} \left( \tau_{m',n'}(t) + \frac{\phi_{m',n'}(t)}{2\pi f_c} \right) \bigg\vert
\end{equation}
\else
\begin{align}
	D_\mathrm{s,\cgls{irs}}(t) = f_c \max_{m, n,m',n'} \bigg\vert &\frac{d}{dt} \left( \tau_{m, n}(t) + \frac{\phi_{m, n}(t)}{2\pi f_c} \right)
	\notag\\
	&- \frac{d}{dt} \left( \tau_{m',n'}(t) + \frac{\phi_{m',n'}(t)}{2\pi f_c} \right) \bigg\vert
\end{align}
\fi
is the Doppler spread across the \cgls{irs}.
Clearly, $D_{\s,0}(t)$ is minimized if
\begin{equation} \label{eq:DSmin}
	\frac{d}{dt} \phi_{m, n}(t) = 2\pi f_c \frac{d}{dt} \left( \tau_0(t) - \tau_{m, n}(t) \right)
\end{equation}
for all $m, n$. Then, $D_{\s,0}(t) = 0$ and
\ifhbonecolumn
\begin{align}
	\MoveEqLeft \frac{d}{dt} \left( \tau_{m, n}(t) + \frac{\phi_{m, n}(t)}{2\pi f_c} \right) - \frac{d}{dt} \left( \tau_{m',n'}(t) + \frac{\phi_{m',n'}(t)}{2\pi f_c} \right)
	\\
	&= 
	\frac{d}{dt} \tau_{m, n}(t) + \frac{d}{dt} \left( \tau_0(t) - \tau_{m, n}(t) \right)
	- \frac{d}{dt} \tau_{m',n'}(t) - \frac{d}{dt} \left( \tau_0(t) - \tau_{m',n'}(t) \right)
	= 0.
\end{align}
\else
\begin{align}
	\MoveEqLeft \frac{d}{dt} \left( \tau_{m, n}(t) + \frac{\phi_{m, n}(t)}{2\pi f_c} \right) - \frac{d}{dt} \left( \tau_{m',n'}(t) + \frac{\phi_{m',n'}(t)}{2\pi f_c} \right)
	\\
	&= 
	\frac{d}{dt} \tau_{m, n}(t) + \frac{d}{dt} \left( \tau_0(t) - \tau_{m, n}(t) \right)
	\notag\\
	&\phantom{={}}- \frac{d}{dt} \tau_{m',n'}(t) - \frac{d}{dt} \left( \tau_0(t) - \tau_{m',n'}(t) \right)
	= 0.
\end{align}
\fi
Hence, \cref{eq:DSmin} also minimizes $D_\mathrm{s,\cgls{irs}}(t)$ and the Doppler spread is zero. From a physical perspective, this choice of $\phi_{m, n}(t)$ compensates for the difference in relative velocities between the direct path and the \cgls{irs} as observed by the receiver. While this approach does not remove the Doppler shift due to the movement of the receiver, it prevents the introduction of additional frequency components in the \cgls{irs} paths.

\subsection{Delay Spread}
The delay spread is the maximum difference in propagation time over all significant transmission paths, i.e.,
$T_\d(t) = \max \{ T_{\d,0}, T_\mathrm{d,\cgls{irs}} \}$
where
\begin{equation}
	T_{\d,0}(t) = \max_{m, n} \left\{  \tau_{m, n}(t) + \frac{\phi_{m, n}(t)}{2\pi f_c} \right\} - \tau_0(t)
\end{equation}
is the delay spread between the \cgls{irs} and direct path,
and
\ifhbonecolumn
\begin{equation}
	T_\mathrm{d,\cgls{irs}}(t) = \max_{m, n} \left\{ \tau_{m, n}(t)+ \frac{\phi_{m, n}(t)}{2\pi f_c} \right\} - \min_{m, n} \left\{ \tau_{m, n}(t)+ \frac{\phi_{m, n}(t)}{2\pi f_c} \right\}
\end{equation}
\else
\begin{multline}
	T_\mathrm{d,\cgls{irs}}(t) = \max_{m, n} \left\{ \tau_{m, n}(t)+ \frac{\phi_{m, n}(t)}{2\pi f_c} \right\} \\- \min_{m, n} \left\{ \tau_{m, n}(t)+ \frac{\phi_{m, n}(t)}{2\pi f_c} \right\}
\end{multline}
\fi
is the delay spread of the \cgls{irs}.

Since $\phi_{m, n}(t) \ge 0$ and due to \cref{eq:delayineq},
\begin{align}
	\min_{m, n} \left\{ \tau_{m, n}(t)+ \frac{\phi_{m, n}(t)}{2\pi f_c} \right\}
	\ge
	\min_{m, n}\left\{ \tau_{m, n}(t) \right\}
	\ge
	\tau_0(t).
\end{align}
Hence,
%\begin{equation}
	$T_\mathrm{d,\cgls{irs}}(t) 
	\le \max_{m, n}\left\{ \tau_{m, n}(t)+ \frac{\phi_{m, n}(t)}{2\pi f_c} \right\} - \tau_0(t) = T_{\d,0}(t)$
%\end{equation}
and the delay spread simplifies to
\begin{equation} \label{eq:Tdopt}
	T_\d(t) = T_{\d,0}(t) = \max_{m, n} \left\{ \tau_{m, n}(t) + \frac{\phi_{m, n}(t)}{2\pi f_c} \right\} - \tau_0(t).
\end{equation}
Thus, the delay spread is an increasing function in $\phi_{m,n}(t)$ and choosing it as small as possible reduces the delay spread. In particular, the unique solution of minimizing $T_\mathrm{d}(t)$ is $\phi_{m,n}(t) = 0$ for all $m, n$ and $t$.

\subsection{Pareto Optimal Phase Shifts} \label{sec:opt}
Our design goal is to simultaneously maximize the received power and minimize delay and Doppler spread, i.e., find a solution to the multi-objective optimization problem
\begin{equation} \label{mop}
	\max_{\forall m, n: \phi_{m, n}(t)} \begin{bmatrix} P_\Rx(t) & -D_\s(t) & -T_\d(t) \end{bmatrix}.
\end{equation}
It can be seen from \cref{eq:Popt,eq:Tdopt} that jointly maximizing $P_\Rx(t)$ and minimizing $T_\d(t)$ is impossible. Hence, \cref{mop} has no single solution but, instead, an infinite number of noninferior solutions \cite{Zadeh1963,Bjornson2014}. A vector $(\phi_{m,n}(t))_{m,n}$ is considered a solution to \cref{mop} if it achieves an objective vector where no component can be improved without worsening at least one other objective.
Such a point is called a Pareto optimal solution and the set of all such vectors is the Pareto optimal solution set of \cref{mop}. Selecting an appropriate solution from this Pareto set is, in general, no trivial task. However, in this case it is easy to make a strong argument for a particular solution.

Clearly, maximum received power is the most important among the objectives in \cref{mop} to ensure good reception as it maximizes the \cgls{snr}. Moreover, as we will show in the sequel, $P_\Rx(t)$ and $-D_\s(t)$ can be maximized simultaneously, and the relative increase in $T_\d(t)$ can be kept small when maximizing the other two objectives. These aspects suggest lexicographic ordering \cite[\S 4.2]{Miettinen1999} as the solution strategy, where the objectives in \cref{mop} are ordered by their absolute importance and maximized successively. In particular, consider the lexicographic order
\begin{equation} \label{eq:lex}
	P_\Rx(t) \succ -D_\s(t) \succ - T_\d(t),
\end{equation}
where ``$\succ$'' stands for ``is more important than'', and let $\mathcal P_\Rx(t)$ be the solution set of maximizing $P_\Rx(t)$, i.e., all solutions that satisfy \cref{eq:Popt}. Then, the lexicographic solution \cref{eq:lex} of \cref{mop} is obtained by first refining $\mathcal P_\Rx(t)$ such that it only contains solutions that minimize $D_\s(t)$ over $\mathcal P_\Rx(t)$ and then selecting a solution from this set that minimizes $T_\d(t)$.

\begin{theorem} \label{thm}
	The lexicographic solution \cref{eq:lex} of \cref{mop} is
	\begin{equation} \label{eq:optphi}
		\phi_{m, n}(t) = 2\pi \mod(f_c (\tau_0(t) - \tau_{m, n}(t)), 1).
	\end{equation}
	It is a Pareto optimal solution of \cref{mop} and
	results in an instantaneous received power
	\begin{equation} \label{eq:optPR}
		P_\Rx(t) = P_\Tx \left| A_0(t) + \sum\nolimits_{m, n} A_{m, n}(t) \right|^2,
	\end{equation}
	Doppler spread $D_\s(t) = 0$, and delay spread
	\begin{align} \label{eq:Tdpareto}
		T_\d(t) = \frac{1}{f_c} \max_{m, n} \lceil f_c (\tau_{m, n}(t) - \tau_0(t)) \rceil.
	\end{align}
\end{theorem}
\begin{IEEEproof}
The received power is maximized for all phase shifts that satisfy \cref{eq:Popt}. The derivative of \cref{eq:Popt} is
\begin{equation}
	\frac{d}{dt} \phi_{m, n}(t) = 2\pi f_c \frac{d}{dt} (\tau_0(t) - \tau_{m, n}(t)) + 2\pi \frac{d}{dt} k_{m, n}(t).
\end{equation}
Except for the last term, this is equivalent to the optimality condition for $\min D_\s(t)$ in \cref{eq:DSmin}.
$k_{m, n}(t)$ is a step function
$k_{m, n}(t) = \sum_{i=1}^n \alpha_i H(t - t_i)$
where $H(t)$ is the Heaviside step function and $\alpha_i \in \{-1, 1\}$. Its derivative is $\frac{d}{dt} k_{m, n}(t) = \sum_{i=1}^n \alpha_i \delta(t - t_i)$ with $\delta(t)$ being the Dirac delta function. Thus, the derivative of $k_{m, n}(t)$ vanishes except for the time instants where $k_{m, n}(t)$ changes its integer value.\footnote{This cannot be prevented since a practical \cgls{irs} is only capable of implementing phase shifts in the order of a few multiples of $2\pi$ \cite[\S V-B]{Tang2019}.} However, since every change in $k_{m, n}(t)$ results in a $2 \pi$ phase shift, it does not lead to discontinuities in the signal and, hence, has no impact on the instantaneous frequency. Therefore, $\frac{d}{dt} k_{m, n}(t)$ can be regarded as effectively zero and \cref{eq:Popt} minimizes the Doppler shift.
From \cref{eq:Tdopt}, the delay spread for \cref{eq:Popt} is
\begin{align}
	&\phantom{={}} T_\d(t) + \tau_0(t) \notag\\
	&= \max_{m, n} \left\{ \tau_{m, n}(t) + \frac{2\pi f_c (\tau_0(t) - \tau_{m, n}(t)) + 2\pi k_{m, n}(t)}{2\pi f_c} \right\} \\
	&= \max_{m, n} \left\{ \tau_{m, n}(t) + \tau_0(t) - \tau_{m, n}(t) + \frac{k_{m, n}(t)}{f_c} \right\}
\end{align}
and, hence, $T_\d(t) = \max_{m, n} \left\{ k_{m, n}(t) \right\} / f_c$.
Thus, $k_{m, n}(t)$ should be chosen as small as possible to minimize the delay spread. Thus, the optimal $k_{m, n}(t)$ is, due to causality and \cref{eq:kcaus}, $k_{m, n}(t) = \lceil f_c (\tau_{m, n}(t) - \tau_0(t)) \rceil$.
Then,
\begin{equation*}
	\begin{aligned}[b]
	\phi_{m, n}(t) &= 2\pi f_c (\tau_0(t) - \tau_{m, n}(t)) + 2\pi \lceil f_c (\tau_{m, n}(t) - \tau_0(t)) \rceil\\
	&= 2\pi \mod(f_c (\tau_0(t) - \tau_{m, n}(t)), 1).
\end{aligned}
%\IEEEQEDhereeqn
\end{equation*}
This solution is Pareto optimal due to \cite[Thm.~4.2.1]{Miettinen1999}.
\end{IEEEproof}

Comparing \cref{eq:Tdpareto} to \cref{eq:Tdopt}, it can be observed that $T_\d(t)$ is increased by at most $\frac{1}{f_c}$ over its physical minimum, which is extremely small compared to the overall delay (spread). Hence, the solution in \cref{thm} jointly optimizes two out of three performance metrics and is very close to the optimal solution of the third. Instead, reversing the lexicographic order in \cref{eq:lex} leads to a slightly smaller delay spread but much larger Doppler spread and no apparent gain of the \cgls{irs}. Changing the order of $P_\Rx(t)$ and $-D_\s(t)$ in \cref{eq:lex} results in the same solution.

\section{Numerical Evaluation} \label{sec:numeval}
\begin{figure}
	\centering
	\begin{tikzpicture}[scale=.8, every node/.append style={font=\footnotesize}]
		\begin{scope}[decoration={markings,mark=at position .5 with {
						\draw[-,white,thick] (-.7pt,0pt) -- (.7pt,0pt);
						\draw[-] (-.7pt,-2.5pt) -- (-.7pt,2.5pt);
						\draw[-] (.7pt,-2.5pt) -- (.7pt,2.5pt);
			}}]
			\draw [->,postaction={decorate}] (0,-2.6)  -- (0,2) node [right] {$y$};
		\end{scope}
		\draw [->] (-1.5,0) -- (2.5,0) node [below] {$x$};
		\draw [dashed] (-1.5,-0.1) -- (2.5,-0.1);

		\begin{scope}
			\clip (current bounding box.north west) rectangle (current bounding box.south east);
			
			\coordinate (tx) at (0,-0.1);
			\draw (tx) -- ++(0,1.5) coordinate (s0);
			%\draw (tx) -- ++(0,-6.3) coordinate (ec) circle [radius=7.8] -- ++(75:7.8) coordinate (s1);
			\draw (tx) ++(0,-6.3) circle [radius=7.8] ++(75:7.8) coordinate (s1);
			\draw (tx) ++(0,-2.5) coordinate (ec) -- (s1);

			\draw[-latex] (ec) +(0,2.1) arc [start angle=90, delta angle=-28.5, radius=2.1];
			\path (ec) ++(75.75:1.7) node {$\alpha(t)$};

			\draw [decorate,decoration={brace,amplitude=3,raise=1}] (tx) -- (s0) node [midway,above,rotate=90,yshift=1] {\SI{1500}{\km}};
			\draw [decorate,decoration={brace,amplitude=3,raise=10,mirror}] (s0) -- (ec) node [midway,left,xshift=-10] {$r_o$};

			\draw [densely dashed] (s1) -- (s1 -| tx);

			%\filldraw (tx) circle [radius=2pt] node [anchor=north west] {};
			\filldraw (s0) circle [radius=2pt] node [anchor=south east] {Rx};
			\draw[-latex] (s0) -- ++(1.75,0) node [pos=0.7,above] {$v$};
		\end{scope}

		\node [draw,anchor=north,xshift=-.5cm] at (current bounding box.north west) {$z = d$};

		\node (t1) [xshift=1.0cm] at (tx -| current bounding box.east) {Earth Surface};
		\node [anchor=west] at (s0 -| t1.west) {LEO};
	\end{tikzpicture}
	\caption{Orbital plane of the satellite and trajectory computation (not to scale).}
	\label{fig:trajectory}
	\vspace{-2ex}
\end{figure}
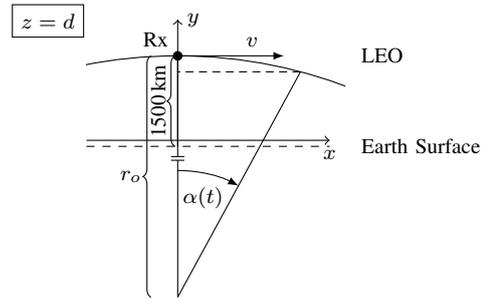

We consider the uplink transmission of a fixed ground terminal, e.g., an \cgls{iot} device, to a satellite in \cgls{leo} at an altitude of \SI{1500}{\km}.
This scenario is described as Deployment-D3 in \cite{TR38.811}. The communication takes place in the S-band at a carrier frequency of $f_c = \SI{2}{\giga\Hz}$ and requires a minimum elevation angle between ground terminal and satellite of \SI{10}{\degree}.
The transmitter is located on the ground with a horizontal distance of \SI{1}{\km} to the \cgls{irs} center, while the \cgls{irs} has an elevation above ground of \SI{100}{\meter}. Hence, $p_\Tx = (0, \SI{-100}{\meter}, \SI{1}{\km})$. The receiver is moving parallel to the $xy$-plane at a horizontal distance $d$ to the \cgls{irs} and in an ideal Keplerian circular orbit. The Earth is assumed as a perfect sphere with radius \SI{6371}{\km}. Thus, the orbital radius of the receiver is $r_o = \SI{7871}{\km}$ and the orbital velocity is $v = \sqrt{\frac{\mu}{r_o}}$ with $\mu$ being Kepler's constant defined as $\mu = \SI{3.986004e5}{\km^3\per\second^2}$ \cite{Ippolito2017}. The computation of the trajectory $p_\Rx(t)$ is illustrated in \cref{fig:trajectory}. In particular, the angle $\alpha(t)$ is obtained from the orbital period $\frac{2\pi r_o}{v}$ as $\alpha(t) = \frac{v}{r_o} t = \sqrt{\frac{\mu}{r_o^3}} t$. Assuming the satellite has position $p_\Rx(0) = (0, \SI{1499.9}{\km}, d)$ at time $t = 0$, its trajectory is computed as
%\begin{equation}
$p_\Rx(t) - p_\Rx(0) = (r_o \sin(\alpha(t)), r_o (\cos(\alpha(t)) - 1), 0)$.
%\end{equation}

The \cgls{irs} has dimensions \SI{18.3 x 12.2}{\meter}, which corresponds to the size of two large US billboards. With an element size of $d_x = d_y = \frac{\lambda_c}{5}$, this amounts to \num{610 x 407} elements that are modeled as
lossless diffuse reflectors ($\mu = 1$) with the planar antenna gain pattern
$G(\theta, \varphi) = \frac{4 \pi}{\lambda_c^2} d_x d_y \cos(\theta)$
for $\theta \in[0, \frac{\pi}{2}]$ and zero otherwise.
With phase shifts as in \cref{eq:optphi}, the channel gain is given as $\frac{P_\Rx(t)}{P_\Tx}$ with $P_\Rx(t)$ as in \cref{eq:optPR}.

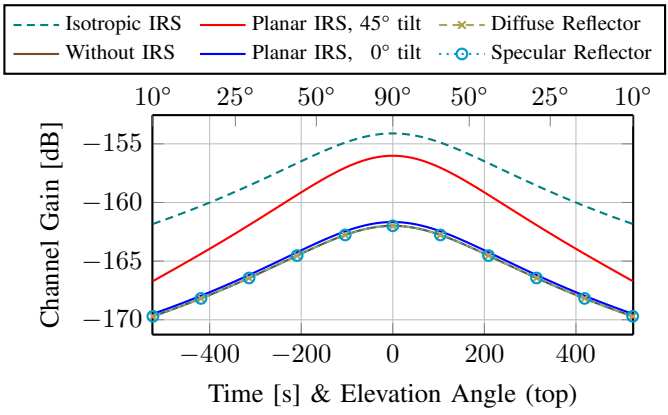
\begin{figure}
\centering
\begin{tikzpicture}[every axis/.append style={
		scale only axis,
			width=.75*\axisdefaultwidth,
			height=.4*\axisdefaultheight,
			thick,
			xmin = -525,
			xmax = 525,
			domain = -525:525,
	}]
	\pgfplotstableread[col sep=comma]{channel.dat}\tbl
	\pgfplotstableread[col sep=comma]{elevation.dat}\elev
	\begin{axis} [
			axis x line*=bottom,
			xlabel={Time [s] \& Elevation Angle (top)},
			ylabel={Channel Gain [dB]},
			ylabel near ticks,
			grid=major,
			%xtick = {0,5,...,30},
			%minor x tick num = 1,
			%ytick = {0, .1, .2, .3},
			%minor y tick num = 3,
			%yminorgrids = true,
			%mark repeat = 2,
			%no markers,
			%ymin = 0,
			%ymax = 10.2,
			legend pos=outer north east,
			legend columns=2,
			transpose legend,
			legend style={font=\footnotesize,at={(current bounding box.north)},yshift=5mm,anchor=south,/tikz/every even column/.append style={column sep=0.15cm}}, %(0.5,1.11)
			legend cell align=left,
			%cycle list name=default,
		]
	
		\addplot[teal,densely dashed] table[y=isotropic IRS] {\tbl};
		\addlegendentry{Isotropic \gls{irs}};

		\addplot[brown!60!black] table[y=no IRS] {\tbl};
		\addlegendentry{Without \gls{irs}};

		\addplot[red] table[y=tilt IRS] {\tbl};
		\addlegendentry{Planar \cgls{irs}, \SI{45}{\degree} tilt};

		\addplot[blue] table[y=planar IRS] {\tbl};
		\addlegendentry{Planar \cgls{irs}, \phantom{0}\SI{0}{\degree} tilt};

		\addplot[ mark=x, mark repeat = 50, densely dashed,yellow!60!black, every mark/.style={solid}] table[y=random] {\tbl};
		\addlegendentry{Diffuse Reflector};

		\addplot[ mark=o, mark repeat = 50, cyan!80!black, dotted, every mark/.style={solid}] table[y=specular] {\tbl};
		\addlegendentry{Specular Reflector};
	\end{axis}
	\begin{axis}[
			axis x line*=top,
			axis y line=none,
			%xlabel={Elevation Angle},
			%xlabel near ticks,
			ymin = -20,
			ymax = -10,
			xtick=data,
			%xticklabels from table={\elev}{Elevation},
			xticklabel = {$\pgfplotstablegetelem{\ticknum}{Elevation}\of{\elev}\pgfmathprintnumber{\pgfplotsretval}\si{\degree}$},
		]
		\addplot table[y=Elevation] {\elev};
	\end{axis}
\end{tikzpicture}
\vspace{-2ex}
\caption{Channel gain over time for one satellite pass. The plots of the scenarios without \cgls{irs} and with reflectors are congruent.}
\label{fig:channel}
\vspace{-2ex}
\end{figure}

Assuming unobstructed view, no atmospheric effects, and isotropic transmit and receive antennas, i.e., $G_\Tx(\theta, \varphi) = G_\Rx(\theta, \varphi) = 1$, the channel gain is displayed in \cref{fig:channel}. As baseline scheme, we compare to the case without \cgls{irs}, i.e., $A_{m, n}(t) = 0$ for all $m, n$ in \cref{eq:optPR}, and to scenarios where the \cgls{irs} is configured to approximate a specular reflector and as a diffuse reflector with phase shifts chosen by Snell's law and uniformly in the interval $[0, 2\pi]$, respectively.
These configurations emulate the behavior of a planar obstacle that reflects the signal in place of the \cgls{irs}.
Further, to obtain an upper bound that serves as a best case deployment, we consider an \cgls{irs} with isotropic elements, i.e., $G(\theta, \varphi) = 1$ for $\theta \in[0, \frac{\pi}{2}]$ and zero otherwise. The number of elements in this case is chosen such that the effective antenna area matches the size of the \cgls{irs}. With an effective area per element of $\frac{\lambda^2}{4\pi}$ this amounts to \num{433 x 288} elements.

\begin{figure}
\centering
\begin{tikzpicture}[every axis/.append style={
		scale only axis,
			width=.75*\axisdefaultwidth,
			height=.40*\axisdefaultheight,
			thick,
			xmin = -525,
			xmax = 525,
			domain = -525:525,
	}]
	\pgfplotstableread[col sep=comma]{gain.dat}\tbl
	\pgfplotstableread[col sep=comma]{elevation.dat}\elev
	\begin{axis} [
			axis x line*=bottom,
			xlabel={Time [s] \& Elevation Angle (top)},
			ylabel={\cgls{irs} Gain [dB]},
			ylabel near ticks,
			grid=major,
			%xtick = {0,5,...,30},
			%minor x tick num = 1,
			%ytick = {0, .1, .2, .3},
			%minor y tick num = 3,
			%yminorgrids = true,
			%mark repeat = 2,
			%no markers,
			%ymin = 0,
			%ymax = 10.2,
			%cycle list name=default,
		]
	
		\addplot[teal,densely dashed] table[y=isotropic IRS] {\tbl};
		\addplot[red] table[y=tilt IRS] {\tbl};
		\addplot[blue] table[y=planar IRS] {\tbl};

		\addplot[ mark=x, mark repeat = 50, densely dashed,yellow!60!black, every mark/.style={solid}] table[y=random] {\tbl};

		\addplot[ mark=o, mark repeat = 50, cyan!80!black, dotted, every mark/.style={solid}] table[y=specular] {\tbl};
	\end{axis}
	\begin{axis}[
			axis x line*=top,
			axis y line=none,
			%xlabel={Elevation Angle},
			%xlabel near ticks,
			ymin = -20,
			ymax = -10,
			xtick=data,
			%xticklabels from table={\elev}{Elevation},
			xticklabel = {$\pgfplotstablegetelem{\ticknum}{Elevation}\of{\elev}\pgfmathprintnumber{\pgfplotsretval}\si{\degree}$},
		]
		\addplot table[y=Elevation] {\elev};
	\end{axis}
\end{tikzpicture}
\vspace{-2ex}
\caption{Gain of using \cgls{irs} over \cgls{los} only communication (legend as in \cref{fig:channel}).}
\label{fig:gain}
\vspace{-2ex}
\end{figure}
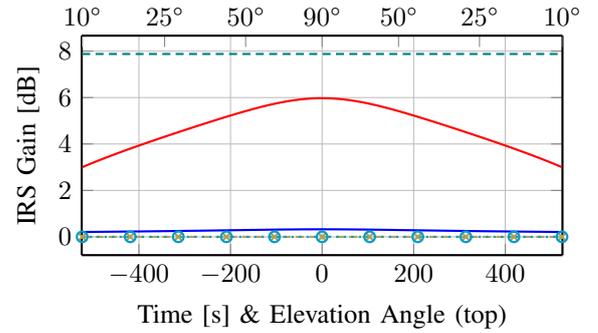

It can be observed from \cref{fig:channel} that the gain of the \cgls{irs} with isotropic elements over the baseline is \SI{7.9}{\dB}. This is also directly displayed in \cref{fig:gain}.
In contrast, the gain of the \cgls{irs} with planar elements is negligible. This is due to the unfavorable angle of the receiver towards the \cgls{irs} which reduces the effective area of the \cgls{irs} to nearly zero. This issue can be avoided by physically rotating the \cgls{irs} towards the sky. In particular, by rotating the $x$-axis by \SI{45}{\degree} and keeping the \cgls{irs} in the $xy$-plane an uptilt of \SI{45}{\degree} is achieved. This results in much better performance as can be observed from \cref{fig:channel,fig:gain}. Depending on the elevation angle, the gain is between \SIrange[range-phrase={ and }]{3}{5.97}{\dB} over the baseline. This amounts to the \cgls{irs}' channel gain being between \SIrange[range-phrase={ and }]{41}{99}{\percent} of the direct channel gain. Hence, the channel over the \cgls{irs} not only results in higher \cgls{snr} but also provides resilience against visibility outages. This comes at the cost of a delay spread that ranges from \SIrange{3.0215}{3.3385}{\micro\second} during the satellite pass, which corresponds to \numrange{6043}{6677} periods of the carrier signal. This indicates a minimum \cgls{cp} below \SI{4}{\micro\second} for \gls{ofdm} transmission whereas 5G supports a maximum \cgls{cp} of \SI{4.7}{\micro\second} \cite{TR21.915}. Observe that the transmission delay is already incorporated in the system model and that the results only depend on the delay spread.
However, rotating a planar reflector in the same way does not result in a noticeable gain over the results in \cref{fig:channel,fig:gain}.
Thus, the observed gains are due to the combination of rotating the \cgls{irs} and configuring it with the Pareto optimal phase shifts from \cref{thm}.

\tikzsetnextfilename{channel}
\tikzpicturedependsonfile{channel.dat}
\tikzpicturedependsonfile{elevation.dat}

\tikzsetnextfilename{gain}
\tikzpicturedependsonfile{gain.dat}
\tikzpicturedependsonfile{elevation.dat}

%elevation angle computation:
%\begin{equation}
%	\frac{\pi}{2} - \arccos\frac{y_\Rx(t) - y_\Tx}{\norm{\vec p_\Rx(t) - \vec p_\Tx}}
%\end{equation}

\section{Conclusions}
We have developed a continuous time model for \cgls{irs}-assisted \cgls{los} communication with a moving receiver. The analysis includes a careful consideration of the time delays. We have shown that the phase shifts can be chosen such that they maximize the received power without incurring any Doppler spread. Moreover, they can be kept within the interval $[0, 2\pi]$ which is in the feasible range of recent \cgls{irs} prototypes \cite{Tang2019}. It also results in the minimum delay spread under meaningful \cgls{irs} operation. In a numerical study, we demonstrate the benefits of \cgls{irs}-assisted \cgls{leo} satellite communication and show that the \cgls{snr} is increased by \SIrange{3}{6}{\dB} for an \cgls{irs} the size of two billboards. This requires rotating the \cgls{irs} such that it has a favorable orientation to the transmitter and receiver. Finding this optimal orientation is left open for future work. Other open topics are the inclusion of statistical channel model to account for atmospheric effects and \cgls{los} outages.

%further work
%\begin{itemize}
%	\item optimize tilt
%	\item include atmospheric effects and obstructions (IRS is always visible, different channel model for Tx to IRS and IRS to Rx)
%	\item different optimization goals if direct path is missing
%\end{itemize}

%\appendices
%\section{Derivative of Norm} \label{app:derivative}
%By the chain rule \cite[Thm.~5.8]{Magnus2007}
%\begin{align}
%	\MoveEqLeft \frac{d}{dt} \Vert \vec x(t) - \vec y(t) \Vert \label{eq:app:deriv}
%	\\
%	&= \left[ D (\Vert \vec x \Vert) (\vec x(t) - \vec y(t)) \right] \left[ \frac{d}{dt} \vec x(t) - \vec y(t) \right]
%\end{align}
%$Df(x)$ is a row vector \cite[Thm.~5.6]{Magnus2007}. From \cite[\S 2.6.1]{Petersen2008} follows $D\Vert\vec x\Vert_2 = \frac{\vec x^T}{\Vert\vec x\Vert_2}$. Thus, the derivative \cref{eq:app:deriv} is
%\begin{equation}
%	\left[ \frac{(\vec x(t) - \vec y(t))^T}{\Vert\vec x(t) - \vec y(t)\Vert} \right] \left[ \frac{d}{dt} \vec x(t) - \vec y(t) \right].
%\end{equation}

\bibliography{IEEEabrv,IEEEtrancfg,references}
\end{document}

%% file: acronyms.tex
\makeglossaries

\newacronym{ofdm}{OFDM}{orthogonal frequency-division multiplexing}
\newacronym{cp}{CP}{cyclic prefix}
\newacronym{los}{LOS}{line-of-sight}
\newacronym{leo}{LEO}{Low Earth Orbit}
\newacronym{iot}{IoT}{Internet of Things}
\newacronym{irs}{IRS}{intelligent reflecting surface}
\newacronym{socp}{SOCP}{second-order cone program}
\newacronym{soc}{SOC}{second-order cone}
\newacronym{dsl}{DSL}{digital subscriber line}
\newacronym{wsee}{WSEE}{weighted sum energy efficiency}
\newacronym{mmwave}{mmWave}{millimeter wave}
\newacronym{dfg}{DFG}{Deutsche Forschungsgemeinschaft}
\newacronym{haec}{HAEC}{Highly Adaptive Energy-Efficient Computing}
\newacronym{hpc}{HPC}{High Performance Computing}
\newacronym{mac}{MAC}{multiple-access channel}
\newacronym{bc}{BC}{broadcast channel}
\newacronym{siso}{SISO}{single-input single-output}
\newacronym{simo}{SIMO}{single-input multiple-output}
\newacronym{miso}{MISO}{multiple-input single-output}
\newacronym{mimo}{MIMO}{multiple-input multiple-output}
\newacronym{af}{AF}{amplify-and-forward}
\newacronym{df}{DF}{decode-and-forward}
\newacronym{cf}{CF}{compress-and-forward}
\newacronym{mwrc}{MWRC}{multi-way relay channel}
\newacronym{dmmwrc}{DM-MWRC}{discrete memoryless multi-way relay channel}
\newacronym{pde}{PDE}{partial data exchange}
\newacronym{fde}{FDE}{full data exchange}
\newacronym{iid}{i.i.d.\@}{independent and identically distributed}
\newacronym{di}{DI} {difference of increasing}
\newacronym{dc}{DC}{difference of convex}
\newacronym{mm}{MM}{mixed monotonic}
\newacronym{mmp}{MMP}{mixed monotonic programming}
\newacronym{awgn}{AWGN}{additive white Gaussian noise}
\newacronym{wgn}{WGN}{white Gaussian noise}
\newacronym{awg}{AWG}{additive white Gaussian}
\newacronym{sic}{SIC}{successive interference cancellation}
\newacronym{snr}{SNR}{signal-to-noise ratio}
\newacronym{sinr}{SINR}{signal to interference plus noise ratio}
\newacronym{inr}{INR}{interference to noise ratio}
\newacronym{zf}{ZF}{zero-forcing}
\newacronym{mrt}{MRT}{maximum ratio transmission}
\newacronym{mmse}{MMSE}{minimum mean square error}
\newacronym{sud}{SUD}{single user decoding}
\newacronym{dof}{DoF}{degrees of freedom}
\newacronym{gdof}{GDoF}{generalized degrees of freedom}
\newacronym{nnc}{NNC}{noisy network coding}
\newacronym{dmn}{DMN}{discrete memoryless network}
\newacronym{csi}{CSI}{channel state information}
\newacronym{pmf}{pmf}{probability mass function}
\newacronym{dmic}{DM-IC}{discrete memoryless interference channel}
\newacronym{ic}{IC}{interference channel}
\newacronym{gic}{GIC}{Gaussian interference channel}
\newacronym{if}{IF}{interference}
\newacronym{ee}{EE}{energy efficiency}
\newacronym{gee}{GEE}{global energy efficiency}
\newacronym{tin}{TIN}{treating interference as noise}
\newacronym{snd}{SND}{simultaneous non-unique decoding}
\newacronym{sd}{SD}{simultaneous decoding}
\newacronym{hk}{HK}{Han-Kobayashi}
\newacronym{rs}{RS}{rate splitting}
\newacronym{rf}{RF}{radio frequency}
\newacronym{pa}{PA}{power amplifier}
\newacronym{lna}{LNA}{low noise amplifier}
\newacronym{lo}{LO}{local oscillator}
\newacronym{adc}{ADC}{analog-to-digital converter}
\newacronym{dac}{DAC}{digital-to-analog converter}
\newacronym{dsp}{DSP}{digital signal processing}
\newacronym{brd}{BRD}{best response dynamics}
\newacronym{br}{BR}{best response}
\newacronym{ne}{NE}{Nash equilibrium}
\newacronym{lhs}{LHS}{left-hand side}
\newacronym{rhs}{RHS}{right-hand side}
\newacronym{ran}{RAN}{radio access network}
\newacronym{qos}{QoS}{Quality of Service}
\newacronym{ngmn}{NGMN}{Next Generation Mobile Networks}
\newacronym{cap}{CAP}{Capacity Adaptation}
\newacronym{bwa}{BW}{Bandwidth Adaptation}
\newacronym{prb}{PRB}{physical resource block}
\newacronym{se}{SE}{spectral efficiency}
\newacronym{tp}{TP}{throughput}
\newacronym{bs}{BS}{base station}
\newacronym{ue}{UE}{user equipment}
\newacronym{mop}{MOP}{multi-objective optimization problem}
\newacronym{gda}{GDA}{generalized Dinkelbach's algorithm}
\newacronym{midcp}{MIDCP}{mixed integer disciplined convex programming}
\newacronym{lp}{LP}{linear program}
\newacronym{brb}{BRB}{branch reduce and bound}
\newacronym{bb}{BB}{branch and bound}
\newacronym{sit}{SIT}{successive incumbent transcending}
\newacronym{oma}{OMA}{orthogonal multiple access}
\newacronym{noma}{NOMA}{non-orthogonal multiple access}
\newacronym{wlog}{w.l.o.g.\@}{without loss of generality}
\newacronym{lsc}{l.s.c.\@}{lower semi-continuous}
\newacronym{usc}{u.s.c.\@}{upper semi-continuous}
\newacronym{kkt}{KKT}{Karush-Kuhn-Tucker}
\newacronym{ptp}{PTP}{point-to-point}

%% file: paper.bbl
\begin{thebibliography}{10}
\providecommand{\url}[1]{#1}
\csname url@samestyle\endcsname
\providecommand{\newblock}{\relax}
\providecommand{\bibinfo}[2]{#2}
\providecommand{\BIBentrySTDinterwordspacing}{\spaceskip=0pt\relax}
\providecommand{\BIBentryALTinterwordstretchfactor}{4}
\providecommand{\BIBentryALTinterwordspacing}{\spaceskip=\fontdimen2\font plus
\BIBentryALTinterwordstretchfactor\fontdimen3\font minus
  \fontdimen4\font\relax}
\providecommand{\BIBforeignlanguage}[2]{{%
\expandafter\ifx\csname l@#1\endcsname\relax
\typeout{** WARNING: IEEEtran.bst: No hyphenation pattern has been}%
\typeout{** loaded for the language `#1'. Using the pattern for}%
\typeout{** the default language instead.}%
\else
\language=\csname l@#1\endcsname
\fi
#2}}
\providecommand{\BIBdecl}{\relax}
\BIBdecl

\bibitem{Liaskos2018}
C.~Liaskos \emph{et~al.}, ``A new wireless communication paradigm through
  software-controlled metasurfaces,'' \emph{{IEEE} Commun. Mag.}, vol.~56,
  no.~9, pp. 162--169, Sep. 2018.

\bibitem{DiRenzo2019}
M.~Di~Renzo \emph{et~al.}, ``Smart radio environments empowered by
  reconfigurable {AI} meta-surfaces: an idea whose time has come,''
  \emph{{EURASIP} J. Wireless Commun. Netw.}, vol. 2019, no.~1, May 2019.

\bibitem{Basar2019}
E.~Basar \emph{et~al.}, ``Wireless communications through reconfigurable
  intelligent surfaces,'' \emph{{IEEE} Access}, vol.~7, pp. 116\,753--116\,773,
  2019.

\bibitem{Wu2020}
Q.~Wu and R.~Zhang, ``Towards smart and reconfigurable environment: Intelligent
  reflecting surface aided wireless network,'' \emph{{IEEE} Commun. Mag.},
  vol.~58, no.~1, pp. 106--112, Jan. 2020.

\bibitem{Wu2019}
------, ``Intelligent reflecting surface enhanced wireless network via joint
  active and passive beamforming,'' \emph{{IEEE} Trans. Wireless Commun.},
  vol.~18, no.~11, pp. 5394--5409, Nov. 2019.

\bibitem{Bjornson2020}
\BIBentryALTinterwordspacing
E.~Björnson and L.~Sanguinetti, ``Power scaling laws and near-field behaviors
  of massive {MIMO} and intelligent reflecting surfaces,'' in review. [Online].
  Available: \url{https://arxiv.org/abs/2002.04960}
\BIBentrySTDinterwordspacing

\bibitem{Tang2019}
\BIBentryALTinterwordspacing
W.~Tang \emph{et~al.}, ``Wireless communications with reconfigurable
  intelligent surface: Path loss modeling and experimental measurement,'' Nov.
  2019. [Online]. Available: \url{https://arxiv.org/abs/1911.05326}
\BIBentrySTDinterwordspacing

\bibitem{Ozdogan2020a}
{\"O}.~{\"O}zdogan, E.~Bj\"{o}rnson, and E.~G. Larsson, ``Intelligent
  reflecting surfaces: Physics, propagation, and pathloss modeling,''
  \emph{{IEEE} Wireless Commun. Lett.}, vol.~9, no.~2, pp. 581--585, May 2020.

\bibitem{Williams2019}
R.~J. Williams, E.~De~Carvalho, and T.~L. Marzetta, ``A communication model for
  large intelligent surfaces,'' in \emph{Proc. IEEE Int. Conf. Commun.
  Workshops (ICC Workshops)}, Dublin, Ireland, 6 2020.

\bibitem{Zadeh1963}
L.~A. Zadeh, ``Optimality and non-scalar-valued performance criteria,''
  \emph{{IEEE} Trans. Autom. Control}, vol.~8, no.~1, pp. 59--60, Jan. 1963.

\bibitem{Bjornson2014}
E.~Bj{\"o}rnson, E.~A. Jorswieck, M.~Debbah, and B.~Ottersten, ``Multiobjective
  signal processing optimization: The way to balance conflicting metrics in
  {5G} systems,'' \emph{{IEEE} Signal Process. Mag.}, vol.~31, no.~6, pp.
  14--23, Nov. 2014.

\bibitem{Bauer2001}
H.~Bauer, \emph{Measure and Integration Theory}.\hskip 1em plus 0.5em minus
  0.4em\relax De Gruyter, 2001.

\bibitem{Miettinen1999}
K.~Miettinen, \emph{Nonlinear Multiobjective Optimization}.\hskip 1em plus
  0.5em minus 0.4em\relax Springer, 1999.

\bibitem{TR38.811}
3GPP, ``Study on {New Radio (NR)} to support non-terrestrial networks,'' Tech.
  Rep. 38.811 V15.2.0, Sep. 2019.

\bibitem{Ippolito2017}
L.~J. Ippolito, Jr., \emph{Satellite Communications Systems Engineering},
  2nd~ed.\hskip 1em plus 0.5em minus 0.4em\relax Wiley, 2017.

\bibitem{TR21.915}
3GPP, ``{5G Rel-15} release description,'' Tech. Rep. TR 21.915 V15.0.0, Oct.
  2019.

\end{thebibliography}
